\def\draft{0} 

\def\lipics{0}

\ifnum\lipics=1
\def\draft{0}
\fi 

\ifnum\lipics=0
\documentclass[11pt,letterpaper]{article}
\usepackage[margin=1in]{geometry}
\usepackage[utf8]{inputenc}
\usepackage{amsthm}
\usepackage{amsthm, amsmath, amssymb, mathtools}
\usepackage{bbm,bm}
\usepackage{graphicx}
\usepackage{color,xcolor}
\usepackage{paralist,enumitem}
\usepackage{mathrsfs}
\usepackage[normalem]{ulem}

\usepackage[colorlinks=true, allcolors=blue]{hyperref}
\usepackage[capitalise,nameinlink]{cleveref}

\else  

\documentclass[a4paper,UKenglish,cleveref, autoref, thm-restate]{lipics-v2021}
\usepackage[utf8]{inputenc}
\usepackage{amsmath, amssymb, mathtools}
\usepackage{bbm,bm}
\usepackage{graphicx}
\usepackage{color,xcolor}
\usepackage{paralist,enumitem}
\usepackage{mathrsfs}
\usepackage[normalem]{ulem}

\fi

\crefname{claim}{Claim}{Claims}

\newcommand{\vnote}[1]{\ifnum\draft=1\textcolor{orange}{[\textbf{Santhoshini:} #1]}\fi}
\newcommand{\mnote}[1]{\ifnum\draft=1\textcolor{red}{[\textbf{Madhu:} #1]}\fi}
\newcommand{\snote}[1]{\ifnum\draft=1\textcolor{red}{[\textbf{Noah:} #1]}\fi}
\newcommand{\sanote}[1]{\ifnum\draft=1\textcolor{red}{[\textbf{Sasha:} #1]}\fi}
\newcommand{\rnote}[1]{\ifnum\draft=1\textcolor{red}{[\textbf{Raghuvansh:} #1]}\fi}
\newcommand{\cnote}[1]{\ifnum\draft=1\textcolor{red}{[\textbf{Chi-Ning:} #1]}\fi}

\usepackage{algorithmicx}
\usepackage{algorithm}
\usepackage[noend]{algpseudocode}
\newcounter{algsubstate}

\algnewcommand\algorithmicinput{\textbf{Input:}}
\algnewcommand\Input{\item[\algorithmicinput]}

\algnewcommand\algorithmicoutput{\textbf{Output:}}
\algnewcommand\Output{\item[\algorithmicoutput]}

\algnewcommand\algorithmicgoal{\textbf{Goal:}}
\algnewcommand\Goal{\item[\algorithmicgoal]}

\newcommand{\Exp}{\mathop{\mathbb{E}}}

\newcommand{\cC}{\mathcal{C}}
\newcommand{\cD}{\mathcal{D}}
\newcommand{\cF}{\mathcal{F}}

\newcommand{\N}{\mathbb{N}}

\newcommand{\R}{\mathbb{R}}
\newcommand{\bern}{\mathsf{Bern}}
\newcommand{\unif}{\mathsf{Unif}}

\newcommand{\mcsp}{\textsf{Max-CSP}}
\newcommand{\mocsp}{\textsf{Max-OCSP}}

\renewcommand{\unif}{\mathsf{Unif}}

\newcommand{\mas}{\textsf{MAS}}

\newcommand{\bias}{\textsf{bias}}
\newcommand{\dbias}{\textsf{d-bias}}
\newcommand{\val}{\mathsf{val}}

\newcommand{\ALG}{\mathbf{ALG}}

\newcommand{\yes}{\textbf{YES}}
\newcommand{\no}{\textbf{NO}}
\newcommand{\mdcut}{\textsf{Max-DICUT}}

\newcommand{\mcut}{\textsf{Max-CUT}}
\newcommand{\maxtwolin}{\textsf{Max-2-LIN}}

\newcommand{\CF}{\mathcal{F}}

\newcommand{\maxf}{\textsf{Max-CSP}(f)}
\newcommand{\maxF}{\textsf{Max-CSP}(\CF)}

\newcommand{\kand}{k\mathsf{AND}}
\newcommand{\maxkand}{\textsf{Max-}\kand}

\newcommand{\maxeksat}{\textsf{Max-Exact-$k$SAT}}
\newcommand{\mkor}{\textsf{Max-$k$OR}}
\newcommand{\maxqcol}{\textsf{Max-$q$Coloring}}
\newcommand{\maxUG}{\textsf{Max-Unique-Games}}

\newcommand{\kgyf}{K_\gamma^Y(f)}
\newcommand{\kbnf}{K_\beta^N(f)}

\newcommand{\sgyf}{S_\gamma^Y(f)}
\newcommand{\sbnf}{S_\beta^N(f)}

\newcommand{\gbmF}{(\gamma,\beta)\textsf{-}\maxF}

\newcommand{\veca}{\mathbf{a}}
\newcommand{\vecb}{\mathbf{b}}

\newcommand{\vecw}{\mathbf{w}}
\newcommand{\vecx}{\mathbf{x}}

\newcommand{\vecsigma}{\boldsymbol{\sigma}}

\newcommand{\vectau}{\boldsymbol{\tau}}

\renewcommand{\tilde}{\widetilde}

\newcommand\eqdef{\stackrel{\mathrm{\small def}}{=}}

\newcommand{\Z}{\mathbb{Z}}

\newcommand{\tv}{\mathrm{tv}}

\DeclarePairedDelimiter{\tvd}{\lVert}{\rVert_{\tv}}

\newcommand{\rhomin}{\rho_{\min}}
\newcommand{\w}{\omega}
\newcommand{\Sym}{\mathrm{Sym}}

\newcommand{\indeg}{\textrm{in-deg}}
\newcommand{\outdeg}{\textrm{out-deg}}

\ifnum\lipics=0 
\usepackage{amsthm}
\usepackage{thmtools,thm-restate}

\numberwithin{equation}{section}
\declaretheoremstyle[bodyfont=\it,qed=\qedsymbol]{noproofstyle}

\declaretheorem[name=Observation,numbered=no]{observation*}

\declaretheorem[numberlike=equation]{theorem}

\declaretheorem[name=Theorem,numbered=no]{theorem*}

\declaretheorem[name=Lemma,numbered=no]{lemma*}

\declaretheorem[name=Corollary,numbered=no]{corollary*}

\declaretheorem[name=Proposition,numbered=no]{proposition*}

\declaretheorem[name=Claim,numbered=no]{claim*}

\declaretheorem[name=Conjecture,numbered=no]{conjecture*}

\declaretheorem[name=Question,numbered=no]{question*}

\declaretheoremstyle[bodyfont=\it]{defstyle}

\declaretheorem[unnumbered,name=Definition,style=defstyle]{definition*}

\declaretheorem[unnumbered,name=Example,style=defstyle]{example*}

\declaretheorem[unnumbered,name=Notation=defstyle]{notation*}

\declaretheorem[unnumbered,name=Construction,style=defstyle]{construction*}

\declaretheoremstyle[]{rmkstyle}

\fi

\ifnum\lipics=1 
\title{Streaming and Sketching Complexity of CSPs: A~survey}
\else 
\title{Streaming and Sketching Complexity of CSPs: A survey\thanks{This paper accompanies an invited talk by the author at ICALP 2022. A version of this paper will appear in the proceedings of the same.}}
\fi

\newcommand{\acktext}{I want to thank Santhoshini Velusamy for introducing me to this wonderful line of research. I want to thank her and all other co-authors --- Chi-Ning Chou, Sasha Golovnev, Noah Singer, Raghuvansh Saxena, Amirbehshad Shahrasbi and Ameya Velingker --- for the collaborations and discussions which informed this survey. They also caught numerous errors and gave suggestions that have hopefully made this survey more readable. I want to thank the organizers of ICALP 2022 for inviting me to write this survey, and to David Woodruff in particular for the additional encouragement and nudges that were crucial to get me going.}

\ifnum\lipics=0
\author{Madhu Sudan\thanks{School of Engineering and Applied Sciences, Harvard University, Cambridge, Massachusetts, USA. Supported in part by a Simons Investigator Award and NSF Awards CCF 1715187 and CCF 2152413. Email: \texttt{madhu@cs.harvard.edu}.}}

\else  

\author{Madhu Sudan}{School of Engineering and Applied Sciences, Harvard University, Cambridge, Massachusetts, USA.}{madhu@cs.harvard.edu}{https://orcid.org/0000-0003-3718-6489}{Supported in part by a Simons Investigator Award and NSF Award CCF 2152413.}

\authorrunning{Madhu Sudan} 

\Copyright{Madhu Sudan} 



\category{Invited Talk} 

\relatedversion{A potentially updated version appears on ECCC as TR 22-065.} 
\relatedversiondetails{Full version}{https://eccc.weizmann.ac.il/report/2022/065/} 

\acknowledgements{\acktext}

\EventEditors{Miko{\l}aj Boja\'{n}czyk, Emanuela Merelli, and David P. Woodruff}
\EventNoEds{3}
\EventLongTitle{49th International Colloquium on Automata, Languages, and Programming (ICALP 2022)}
\EventShortTitle{ICALP 2022}
\EventAcronym{ICALP}
\EventYear{2022}
\EventDate{July 4--8, 2022}
\EventLocation{Paris, France}
\EventLogo{}
\SeriesVolume{229}
\ArticleNo{5}

\fi 

\date{May 5, 2022}

\begin{document}

\maketitle

\begin{abstract}
    In this survey we describe progress over the last decade or so in understanding the complexity of solving constraint satisfaction problems (CSPs) approximately in the streaming and sketching models of computation. After surveying some of the results we give some sketches of the proofs and in particular try to explain why there is a tight dichotomy result for sketching algorithms working in subpolynomial space regime.
\end{abstract}

\ifnum\lipics=0 \tableofcontents \newpage \fi

\section{Introduction}

In this article we survey the current state of knowledge on the {\em approximability} of {\em constraint satisfaction problems (CSPs)} using small space {\em streaming} and {\em sketching} algorithms. We start by reviewing the definitions below before turning to the survey of results.

\section{CSPs: What and Why}

CSPs form an infinite class of optimization problems where the goal is to assign $n$ variables values from a finite set while maximizing the number of constraints that can be satisfied, where each constraint looks locally at the assignment of a few variables to determine if it is satisfied on not.
Different problems in the class differ based on which type of constraints are allowed. Different instances of the problem arise by applying the constraints to different subsets (or subsequences) of variables. Algorithms aim to compute, or approximate, the maximum, over all assignments, of the fraction of constraints that can be simultaneously satisfied. Resource restrictions on the algorithm (time, space, number of passes in the streaming setting) as well as the type of constraints allow to determine the level of approximability that is feasible. In this survey we describe our knowledge of the approximability of CSPs when restricted to streaming and sketching algorithms with limited space. 

We start by describing CSPs more formally. For positive integer $n$ we use $[n]$ to denote the set $\{1,\ldots,n\}$ and $\Z_n$ to denote the set $\{0,\ldots,n-1\}$. A CSP problem is described by positive integers $k$, $q$ and a family of functions $\cF \subseteq \{f:\Z_q^k \to \{0,1\}\}$. Since $k,q$ are implicit in $\cF$, we refer to this problem as $\maxF$. Given variables $X_1,\ldots,X_n$, an assignment to the variables is a sequence $\veca = (a_1,\ldots,a_n) \in \Z_q^n$. A constraint $C$ on these variables is given by a pair $(f,(j_1,\ldots,j_k))$ where the first element of the pair $f \in \cF$ is the choice of the type of constraint and the second element is a sequence of $k$ {\em distinct} indices with $j_i,\ldots,j_k \in [n]$.
An assignment $\veca$ satisfies $C = (f,(j_1,\ldots,j_k))$ if and only if $f(a_{j_1},\ldots,a_{j_k})=1$.
We use $C(\veca)$ to denote the quantity $f(X_{j_1},\ldots,X_{j_k})$. An instance of $\maxF$ on $n$ variables and $m$ constraints is given by $\Psi = (C_1,\ldots,C_m)$ with $C_i$ being a constraint on $X_1,\ldots,X_n$ for every $i \in [m]$. Given a assignment $\veca$ to the variables, the value on the instance $\Psi$ at $\veca$, denoted $\val_{\Psi}(\veca)$, is the quantity $\frac{1}{m}\sum_{i\in[m]} C_i(\veca)$, i.e., the value is the fraction of constraints of $\Psi$ that are satisfied by $\veca$. The value of the instance $\val_{\Psi}$ is defined to be the maximum value over all assignments, i.e., $\val_{\Psi} = \max_{\veca \in [q]^n} \{\val_{\Psi}(\veca)\}$. The goal of CSP optimization algorithms is to compute, or approximate, $\val_\Psi$ given $\Psi$.\footnote{Throughout this paper we assume that $\cF$ does not include the all $0$ function. Such a function corresponds to placing constraints that are never satisfiable. Inclusion of such constraints in the family does not change the complexity of any of the tasks we consider since these constraints are easy to ignore.} 

\medskip

\noindent {\bf Example:} We illustrate the definition with the example of the $\mcut$ problem, where given an undirected graph on vertex set $[n]$, the goal is to find a ``cut'' $S \sqcup \overline{S} = [n]$ that maximizes the number of edges crossing the cut (i.e, with one endpoint each in $S$ and $\overline{S}$). This problem is captured by $q=k=2$ and the family $\cF = \{\oplus\}$ where $\oplus(u,v) = u+v \pmod 2$. 
\ifnum\lipics=0 
For instance if the input graph for $\mcut$ is the 5-cycle, then the input instance $\Psi_5$ to $\maxF$ will have $5$ variables (corresponding to the vertices) and 5 constraints $C_1,\ldots,C_5$ (corresponding to the edges) with $C_i = (\oplus, (i, i+1))$ for $i \in [4]$ and $C_5 = (\oplus, (5,1))$. An assignment $\veca \in \Z_2^5$ can be equated with the cut $S_{\veca} = \{i | a_i = 1\}$ and it can be verified that $\val_{\Psi_5}(\veca)$ equals the fraction of edges cut by $S_{\veca}$. 
\fi 

Note that in the example above, we could get positive integer weighted graphs also since there is no requirement that the constraints themselves be distinct. For simplicity we will assume the length on the stream is polynomial in $n$ so that the weights are also polynomial, though as we point out later this restriction does not alter the complexity of the approximation problems much. But the graphs will not have any self-loops due to the ``restriction'' that the variables in a constraint have to be distinct. If one wished to consider the $\mcut$ problem where self-loops are also allowed (though in the case of $\mcut$ this would make no sense - since a self-loop can never be cut), then one could consider instances $\mcsp(\cF')$ where $\cF' = \{\oplus,F\}$ and $F(u,v) = \oplus(u,u)$ for all $u,v$. (So $v$ is just a dummy variable and $u$ is a variable which supplies both arguments to the cut function $\oplus$.) Thus while the requirement that the variables are distinct may appear as a restriction, it is not: For every family $\cF$ there exists a family $\cF'$ such that $\mcsp(\cF')$ captures the $\maxF$ problem where variables are allowed to repeat. 

We also remark that in some prior works in the Boolean setting, i.e., when $q=2$, constraints may be applied to ``literals'', rather than ``variables''. We refer to these results as applying to Boolean CSPs over literals. In our setting we apply constraints only to variables. Again, our setting is more general than the setting of Boolean CSPs over literals in that for every family $\cF \subseteq \{f:\Z_2^k \to \{0,1\}\}$ there is a family $\cF'$ such that CSP with constraints from $\cF$ applied to literals is the same problem as $\mcsp(\cF')$. For example consider the $\maxtwolin$ problem, i.e., the CSP whose constraints are given by linear equations modulo $2$ on two distinct variables. $\maxtwolin$ is the Boolean CSP over literals over the family $\cF = \{\oplus\}$. But if we let $\cF' = \{\oplus, \overline{\oplus}\}$ where $\overline{\oplus}(u,v) = u+v+1 \pmod 2$, then $\maxtwolin = \mcsp(\cF')$. On the other hand $\maxF$ is the $\mcut$ problem which can not be expressed as a Boolean CSP over literals. Thus our setting is strictly richer in expressibility.

\subsection{Why study CSPs?}
Before going on to giving more precise descriptions of the approximation versions of CSPs and models of streaming algorithms, we digress to comment on why study CSPs at all. 

CSPs do capture a host of natural optimization problems: Some familiar names of problems include the Maximum Cut problem in (undirected) graphs, the Maximum Dicut problem in directed graphs, the Maximum $q$-colorability problem in graphs, the Unique Games problem, etc. Each one of these problems on their own right has probably been the topic of multiple papers, and the umbrella of CSPs unifies their study.
That being said this reason on its own is not as compelling as some of the other reasons we describe next --- after all the study of CSPs does exclude many other natural optimization problems including problems based on connectivity in graphs such as flow maximization or congestion minimization. It also excludes global considerations such as balanced cuts or sparse cuts; and of course there are a host of non-graph-theoretic problems. 

To this author, the real reason to study CSPs is that they tend to allow for finite classification. The first such result dates back to Schaefer~\cite{Sch78} who studied the satisfiability of Boolean CSPS and showed that they exhibit a dichotomy. Feder and Vardi~\cite{FederV} explored the expressibility of different logics and arrived at a morally ``broadest'' set of problems (``Monotone Monadic SNP'') that could potentially exhibit a dichotomy, and showed that this set of problems was essentially equivalent to CSPs over arbitrary finite alphabets. They posed the dichotomy of this class as an open question which was eventually resolved by Bulatov~\cite{Bul17} and Zhuk~\cite{Zhu20} after many years of sustained attack.  Subsequent works extended such classification quests to other classes of problems including optimization and counting. (See Creignou, Khanna and Sudan~\cite{CKS01} for some of the early lines of work.) Many of these bodies are extensive, see e.g. the recent monograph by Cai and Chen~\cite{CaiChen} and references therein for vast explorations of counting problems. In optimization and approximation the work of Raghavendra~\cite{Rag08} gives a fine dichotomy, under the ``Unique Games Conjecture'', that inspires some of the streaming work we describe in this survey.

Finite classifications are interesting in that they highlight the generality of some algorithms. Even the  weak classification of the approximability of CSPs pointed to the general utility of the randomized rounding and Max Flow algorithms~\cite{KSTW01}. The sharp characterizations in \cite{Rag08} point to the power of semidefinite programming and specifically to the sum-of-squares framework of algorithms. One could also ask similar questions in the context of streaming: The dichotomy work presented in this survey does highlight the role of norm estimation algorithms in streaming optimization. Other algorithms might emerge with further exploration.

Finite classifications also point to interesting phenomena. For instance in the context of polynomial time approximability most natural problems have approximability in well separated bands of functions: constant factor approximations, polylogarithmic approximations, and polynomial approximations are common whereas very few have intermediate approximability, say to within a factor of $2^{\sqrt{\log n}}$. A finite classification implies this phenomenon - the entire infinite class of functions only shows finitely many distinct behaviors. A similar phenomenon again seems to occur with streaming algorithms --- many problems have polylogarithmic space approximation algorithms while others require polynomially growing space. Intermediate complexity is rare. CSPs again seems to validate this separation, at least in the context of sketching algorithms, as we will see in the rest of this survey. 

\section{Preliminaries: Approximation and Streaming}

We formalize some basic notions related to approximation problems and streaming algorithms. While a reader familiar with the notion might skip ahead we recommend they make sure they understand the notions (and notations) of: (i) trivial approximation algorithms, (ii) gapped optimization problems and the notation: $\gbmF$ (iii) approximability and approximation-resistance and (iv) sketching algorithms.

\subsection{Approximating CSPs}

Since solving $\maxF$ exactly can be quite hard for most $\cF$ we turn often to algorithms that produce approximate solutions. We discuss some basic definitions regarding these in this section. All definitions restrict algorithms to come from some resource-bounded class $\cC$. While we defer the discussion of the specific classes considered to later sections, here we consider definitions for a generic such class $\cC$. 

The most common notion of approximation is an $\alpha$-approximation algorithm for some $\alpha \in [0,1]$. An $\alpha$-approximation algorithm $\ALG$ for $\maxF$ is one that for every instance $\Psi$ outputs a value $\ALG(\Psi)$ satisfying $\alpha \cdot \val_{\Psi} \leq \ALG(\Psi) \leq \val_\Psi$.\footnote{Note that a small space streaming algorithm has very little hope of outputting a near-optimal solution which might take $\Omega(n)$ space to represent. So we require our algorithms only to output the value achieved by the optimal, or approximately-optimal, solution.} We say $\maxF$ is $\alpha$-approximable in $\cC$ if there exists $\ALG \in \cC$ that is an $\alpha$-approximation algorithm for $\maxF$.

A more refined notion of approximation that is more common in the literature proving non-existence of algorithms is associated with gapped problems. Given $0 \leq \beta < \gamma \leq 1$, we say that an algorithm $\ALG$ solves the ``$(\gamma,\beta)$-approximation version of $\maxF$'', abbreviated $\gbmF$, if the following two conditions hold: (1) For every $\Psi$ such that $\val_\Psi \geq \gamma$, we have $\ALG(\Psi) = 1$ and (2) For every $\Psi$ such that $\val_\Psi \leq \beta$, we have $\ALG(\Psi) = 0$. We say that $\gbmF$ is solvable in $\cC$ if there exists an $\ALG \in \cC$ solving $\gbmF$. 

Assuming $\cC$ satisfies mild closure properties the latter notion roughly captures $\alpha$-approximability precisely, while giving more detailed information. To see some flavor of the translation between the two notions, suppose $\maxF$ is $\alpha$-approximated by $\ALG$. Now consider a pair $\gamma,\beta$ with $\beta < \alpha \gamma$. Then the algorithm $\ALG'$ that outputs $\ALG'(\Psi) =1$ if $\ALG(\Psi) > \beta$ and $0$ otherwise solves $\gbmF$. And if $\cC$ satisfies the closure property that $\ALG' \in \cC$ whenever $\ALG \in \cC$ then it follows that $\alpha$-approximability in $\cC$ implies $\gbmF$ is solvable in $\cC$ for every $\beta < \alpha \gamma$. A rough converse is also true, again assuming some (this time more stronger) closure properties of $\cC$ that we leave unspecified: If for some $\alpha$ we have that for every $\gamma \in [0,1]$, the $\gbmF$ is solvable in $\cC$ for $\beta = \alpha \gamma$, then for every $\epsilon > 0$ we have that $\maxF$ is $\alpha-\epsilon$ approximable in $\cC$. (See \cite[Proposition 2.21]{CGSV21-finite} for a detailed statement and proof.) 

The discussion above explains why the study of $\gbmF$ (for every $\gamma, \beta$ and $\cF$) is at least as rich as the study of $\alpha$-approximability of $\maxF$. But it can provide more detailed information. For instance researchers are often interested in approximating the value on satisfiable, or nearly satisfiable, instances. (See for instance, \cite{DalmauK13,BartoK12,KunOTYZ12} for such works in the setting of $\cC$ being all polynomial time algorithms.)
We can understand these corner cases by focussing on $\gamma=1$ or $\gamma \to 1$ and exploring the maximal $\beta$ such that $\gbmF$ is solvable in $\cC$. For instance recent results show that for every $\beta < 1$, $(1,\beta)$-$\mdcut$ is solvable in $\cC$ when $\cC$ is the class of polylog space bounded sketching algorithms --- a result that is not captured by the single parameter approximability of the problem.

Before concluding we also highlight what is a ``non-trivial'' approximation. For families $\cF$ where every constraint has at least one satisfying assignment this notion is quite simple. We say that an algorithm that outputs a constant (independent of the input $\Psi$) is a trivial algorithm. Note that trivial algorithms are still legitimate approximation algorithms. For instance the algorithm that always outputs $1/2$ is a $1/2$-approximation for $\mcut$ --- this is so since no instance $\Psi$ of $\mcut$ has $\val_\Psi < 1/2$. We say that an approximation ratio is non-trivial if it is not achieved by a trivial algorithm. Similarly we say that a $(\gamma,\beta)$-approximation is non-trivial if $\gamma < 1$ and there exists an instance $\Psi$ with $\val_\Psi \leq \beta$. To quantify this notion of non-triviality we define $\rhomin(\cF)$ to be $\inf_{\Psi {\rm~instance~of~}\maxF} \{\val_{\Psi}\}$. We say that a $\maxF$ is {\em approximation resistant} to $\cC$ if for every $\alpha > \rhomin(\cF)$ no algorithm $\ALG\in\cC$ is an $\alpha$-approximation to $\maxF$. Equivalently for every $\rhomin(\cF) < \beta < \gamma < 1$ it is the case that $\gbmF$ is not solvable in $\cC$. We say $\maxF$ is {\em approximable} in $\cC$ if it is not {\em approximation resistant} to $\cC$. 

In what follows we will describe works exploring the various approximation factors achievable for $\maxF$ for different $\cF$ with streaming algorithms that have bounded space. We shall also explore some restrictions of streaming algorithms known as sketching algorithms. We introduce these terms below.

\subsection{Streaming and Sketching algorithms}

We consider the approximability of $\maxF$ when the input instance $\Psi = (C_1,\ldots,C_m)$ is presented as sequence of constraints to the approximating algorithm. The algorithm is restricted in the amount of space it is provided. We allow the algorithm to be randomized: in all upper bounds the algorithm will be expected to generate and store any randomness in the restricted space it is given, while in the lower bounds we will rule out algorithms that are a distribution over deterministic algorithms (and so strictly more general). 

Formally a space $s(n)$-streaming algorithm for $\gbmF$ on $n$ variables is given by a pair of functions $(\tau,\nu)$ where $\Gamma:\{0,1\}^{s(n)} \times \Lambda_n \to \{0,1\}^{s(n)}$ is the state transition function and output function $\nu:\{0,1\}^{s(n)} \to \{0,1\}^{s(n)}$, where $\Lambda_n=\Lambda_n(\cF)$ denotes the set of all possible constraints of $\maxF$ on $n$ variables.
On input a stream $\vecsigma = (C_1,\ldots,C_m) \in \Lambda_n^m$ the algorithm first computes the state $S(\vecsigma) = S_m$ where $S_i = \Gamma(S_{i-1},C_i)$ for $i \in [m]$ and $S_0 = 0$ in the deterministic case. It then outputs $\nu(S(\vecsigma))$.  A randomized streaming algorithm is the same except that now the initial state $S_0$ is distributed uniformly randomly over $\{0,1\}^{s(n)}$. 

In this paper we will also focus on a restriction of streaming algorithms known as sketching algorithms. To define this notion consider the concatenation of two streams $\vecsigma \circ \vectau$. By definition of a streaming algorithm the final state $S(\vecsigma \circ \vectau)$ can be determined from $S(\vecsigma)$ and $\vectau$. A sketching algorithm is one that is restricted even further in that $S(\vecsigma \circ \vectau)$ can be determined from $S(\vecsigma)$ and $S(\vectau)$, i.e., there exists a composer function $C:\{0,1\}^{s(n)} \times \{0,1\}^{s(n)} \to \{0,1\}^{s(n)}$ such that for every $\vecsigma,\vectau \in \Lambda_n^*$, we have $S(\vecsigma \circ \vectau)=C(S(\vecsigma),S(\vectau))$.


Most common sketching algorithms are obtained from so called ``linear-sketching algorithms'' where $C \in \Gamma_n$ is viewed as a vector in $v \in \R^N$ for some large $N$, and a stream $(C_1,\ldots,C_m)$ represents the sum of the $m$ corresponding vectors $v_1 + \cdots + v_m$. The sketch of a vector $v$ is given by $Av$ where $A \in \R^{N \times s}$ projects $v$ down to some low-dimensional subspace. Ignoring bit precision issues this compresses large $N$ dimensional inputs into small $s$ dimensional sketches that end up giving significant information about the original input, surprisingly often. It is easy to see that such linear sketching algorithms indeed satisfy the definition of sketching.

In the rest of this article we describe the surprising effectiveness of sketching algorithms in approximating $\maxF$. We also describe matching lower bounds for sketching algorithms that often generalize also to give streaming lower bounds.

\section{Results on Streaming CSPs}

Prior to 2010, despite extensive work on streaming algorithms and lower bounds for other problems, there were no works covering CSPs. This was even noted at a workshop at Bertinoro in 2011~\cite{IMNO11}. 

\paragraph*{Lower bounds for $\mcut$.}
The first works focussed on lower bounds for the $\mcut$ problem in independent works by Kogan and Krauthgamer~\cite{KK15} and Khanna, Kapralov and Sudan~\cite{KKS15}. The former showed that there existed $\alpha < 1$ such that $\alpha$-approximating $\mcut$ requires $\Omega(\sqrt{n})$-space. The latter showed the tighter result showing that for every $\alpha > 1/2$, $\alpha$-approximating $\mcut$ requires $\Omega(\sqrt{n})$-space in the streaming setting. In other words $\mcut$ is approximation resistant to $o(\sqrt{n})$ space streaming algorithms. Subsequent works focussed on the space complexity and pushed it higher. Khanna, Kapralov, Sudan and Velingker~\cite{KKSV17} pushed the space requirement up to linear at the cost of a weaker approximation; specifically they showed that there exists $\alpha < 1$ such that $\alpha$-approximating $\mcut$ requires $\Omega(n)$-space. Finally, in a tour-de-force work, Kapralov and Krachun~\cite{KK19} settled the approximability of $\mcut$ essentially completely by showing it is approximation resistant to $o(n)$ space streaming algorithms. 

An aside is in order here: The input to a $\maxF$ has length $O(m \log n)$ and even if we forbid repeated constraints $m$ can be as large as $n^k$. So, a priori one could imagine space complexities of streaming algorithms being much higher than $O(n)$. But a folklore observation shows that it suffices for the streaming algorithm to maintain a random sample of $O(n/\epsilon^2)$ constraints and the optimum value on the sampled constraints is a $(1-\epsilon)$ approximation to the optimum value on the input instance.\footnote{This observation also relies on the fact that the value of every instance is bounded away from $0$, which in turn relies on the fact that $0$ is not a constraint function in $\cF$.} Since this sample of constraints takes only $O(n \log n)$ space to store and the optimal value for this sample can be computed in $O(n)$ space (though using exponential time), it follows that $O(n \log n)$ space suffices to get $\alpha$-approximation for every $\maxF$ problem for every $\alpha < 1$. In particular, returning to the $\mcut$ problem, the space lower bound from \cite{KK19} is optimal to within a logarithmic factor.

\paragraph*{Upper bounds for $\mdcut$.}
While the lower bounds for $\mcut$ are technically hard (especially \cite{KK19}) arguably these results are not very surprising: A streaming algorithm with limited  (say polylogarithmic) space seems hardly capable of understanding the global structure imposed by the many different constraints and understanding how well they can be satisfied simultaneously. Indeed the lower bounds of \cite{KKS15} show that $o(\sqrt{n})$ space streaming algorithms can not distinguish random graphs from random bipartite graphs with a planted bipartition. In view of such limited power to understand the global structure it would not have surprised some researchers (notably this author) if every $\maxF$ problem had turned out to be approximation resistant to $o(n)$-space streaming algorithms. In other words --- it was conceivable in 2015 that there were no non-trivial streaming algorithms for CSPs. 

A striking paper of Guruswami, Velingker and Velusamy~\cite{GVV17} changed the picture by giving an elegant and simple algorithm for approximating $\mdcut$, the problem whose input is a directed graph on vertex set $[n]$ and the goal is to find a cut $S \sqcup \overline{S}$ that maximized the number of edges going from $S$ to $\overline{S}$. (This problem is expressible as $\maxF$ for $\cF = \{u \wedge \overline{v}\}$.) The ``trivial'' approximability of this problem is $1/4$, but \cite{GVV17} gave a non-trivial $0.4$-approximation algorithm for this problem using polylogarithmic space.\footnote{Throughout this article we will not spell out the exponent in polylogarithmic terms though of course the original papers give more detailed answers.} The key insight to their algorithm is that one can estimate some non-trivial global information about the input by appealing to norm estimation algorithms that have been well explored in the sketching community. In particular their work relies on algorithms for estimating the $\ell_1$ norm of a vector in the ``turnstile'' model which go back to the work of Indyk~\cite{Ind06}. 
As we will discuss later, this algorithm can be generalized arguably quite surprisingly to many other CSPs. 

\paragraph*{Tight bounds and classification of Boolean binary CSPs.}
While the lower bound for $\mcut$ is obviously tight, the $\mdcut$ approximability of $.4$ from \cite{GVV17} was not known to be tight. From the $1/2+\epsilon$-inapproximability of $\mcut$ one can deduce a $(1/2+\epsilon)$-inapproximability, for every $\epsilon>0$, for $\mdcut$ as well 
(by a reduction which maps every edge from an instance of $\mcut$ to a pair of directed edges between the same vertices). Neither the algorithm nor the analysis appear tight. Indeed in a subsequent work, Chou, Golovnev and Velusamy~\cite{CGV20}  managed to improve both the algorithm and the lower bound to get a tight approximability of $4/9$ for $\mdcut$. Specifically they give a polylog space algorithm achieving this approximation ratio and also prove that no streaming algorithm with $o(\sqrt{n})$ space can do better! This tight result for $\mdcut$ may appear accidental, but \cite{CGV20} go further and classify the approximability of every Boolean (i.e., with $q=2$) CSP on literals on binary constraints (i.e., $k=2$). In doing so their work points to some remarkable phenomena: For every $\alpha\in[0,1]$, every CSP in the finite, but nevertheless diverse, class they consider either is $\alpha$-approximable in polylogarithmic space, or is not $\alpha-\epsilon$ approximable (for every $\epsilon >0$) with $o(\sqrt{n})$ space. And in all cases the approximation algorithm uses the $\ell_1$-norm approximator in a manner similar to \cite{GVV17}. Together these results suggest a broader phenomenon explored and somewhat confirmed in the further work reported next.

\paragraph*{Sketching complexity of CSPs.}

In joint work with Chou, Golovnev and Velusamy~\cite{CGSV21-finite} we give a dichotomy result for all $\gbmF$ (i.e., for every $k,q,\cF$ and every $\gamma,\beta \in [0,1]$) for $o(\sqrt{n})$-space sketching algorithms, as described below.

\begin{theorem}[{\protect \cite[Theorem 1.1]{CGSV21-finite}}]\label{thm:cgsv}
For every $q,k \in \N$, $0 \leq \beta < \gamma \leq 1$, and every $\cF \subseteq \{\Z_q^k \to \{0,1\}\}$, one of the following two conditions holds:
Either $\gbmF$ can be solved by a polylogarithmic space sketching algorithm, or for every $\epsilon > 0$, every sketching algorithm for $(\gamma-\epsilon,\beta+\epsilon)$-$\maxF$ requires $\Omega(\sqrt{n})$-space. 
Furthermore there is a polynomial space algorithm that decides, given $\gamma,\beta$ and $\cF$, which of the
two conditions holds.
\end{theorem}

A corollary to approximation resistance is the following: For every $\cF$, either $\maxF$ is approximable by a polylogarithmic space sketching algorithm, or it is approximation resistant to $o(\sqrt{n})$-space sketching algorithms.\footnote{This corollary is not immediate from the theorem statement, but uses some additional aspects of the proof. See~\cite[Theorem 2.14]{CGSV21-finite} for details.}
In the special case of $k=q=2$ the lower bound above extends beyond sketching algorithms to all streaming algorithms (\cite[Theorem 1.3]{CGSV21-finite}). Put together these results subsume all previous works with the exception of the linear space lower bound for $\mcut$ from \cite{KK19}. Even in the case of $k=q=2$ the dichotomy is more detailed than the one in \cite{CGV20} in that it covers all CSPs, not just CSPs on literals, and it also talks about the solvability of all $\gbmF$ and not only the best approximation ratio. For example the results show that for every sufficiently small $\epsilon>0$, $(1-\epsilon,1-2\epsilon)$-$\mdcut$ is solvable by a polylogarithmic space sketching algorithm while $(1-\epsilon,1-2\epsilon+\delta)$-$\mdcut$ requires $\Omega(\sqrt{n})$ space for every streaming algorithm for every $\delta > 0$. In particular it asserts that nearly satisfiable instances are detectable by small space sketching algorithms. 

The sketching algorithms used for the positive result in \cref{thm:cgsv} builds on the algorithm of \cite{CGV20}, which we refer to as a ``bias-based algorithm'' here. We will discuss that algorithm further later, but highlight one major difference. Rather than appealing to $\ell_1$-norm estimation algorithms, the new algorithm appeals to a matrix norm estimation algorithm, this time from the work of Andoni, Krauthgamer and Onak~\cite{AKO11}. (Roughly
the $\ell_1$ norm given by $\|(x_1,\ldots,x_n)\|_1 = \max_{b_1,\ldots,b_n \in \{-1,+1\}} \sum_{i=1} b_i x_i$ optimizes over a Boolean domain.\snote{Isn't it standard to denote this $\|(x_1,\ldots,x_n)\|_1$?} The matrix norm estimators allow us to optimize some problems over $q$-ary domains.)

While the theorem holds out the possibility that there are non-trivial approximation algorithms for (infinitely) many CSPs, this is not immediate from the theorem statement due to the lack of ``explicitness'' of the classification. Specifically there is no simple relationship that says given $\cF$ what range of $\gamma$ and $\beta$ are ``easy'' (i.e., solvable in polylog space) and which ones are not. This is unfortunately inevitable. As $\cF$ gets more complex the relationships do seem to get more complex. The results of \cite{CGSV21-finite} show that $\gamma$ and $\beta$ are determined by optimizing some $O(q^k)$ real variable linear function over the reals subject to some degree $k$ polynomial constraints. Even in the case of $\mdcut$ this leads to some degree $2$ polynomials in $\gamma$ and $\beta$ that determine the complexity. (See \cite[Example 1, Pages 21-23]{CGSV21-finite} for more details.) Nevertheless the conditions can be analyzed computationally, and in particular using the quantified theory of reals (using only the existential theory does not seem to suffice) to understand the complexity of $\gbmF$ for any given $\gamma,\beta,\cF$. 

Remarkably some subsequent work has managed to extract explicit results, even for infinite families of functions, by exploring the decision conditions arising from the proof of \cref{thm:cgsv}. For instance, Boyland, Hwang, Prasad, Singer and Velusamy~\cite{BHPSV}, analyze the approximability of $\maxkand$ for every $k \in \N$ --- the problem where constraints are the conjunctions of $k$-literals --- and give an exact expression for the approximation ratio of $\maxkand$. (They show $\maxkand$ is approximable to within a factor that roughly looks like $2^{-(k-1)}(1- O(1/k))$ - see \cite{BHPSV} for an exact expression.) In particular this gives an infinite subfamily of CSPs that is non-trivially approximable by the algorithm from \cite{CGSV21-finite}. \cite{BHPSV} also pin down the approximability of some other symmetric functions. Another work, by Chou, Golovnev, Shahrasbi, Sudan and Velusamy~\cite{CGShSV}, also analyzes the sketching approximability of some linear threshold functions, giving some infinite families that are approximation-resistant to $o(\sqrt{n})$-space sketching algorithms and other infinite families that are approximable by polylog space sketching algorithms.  

\paragraph*{Streaming Lower Bounds.}

While the classification essentially only rules out sketching algorithms using $o(\sqrt{n})$-space for the hard problems, for a broad class of problems it even rules out non-trivial streaming algorithms. In fact for all problems it pins down the polylogarithmic space approximability to within a factor of $q$ --- we expand on this later below, but first speak about broad classes of approximation resistant problems. We start with some definitions.

We say that a distribution $\cD$ on $\Z_q^k$ is one-wise independent if for every $i\in[k]$ we have that when $X = (X_1,\ldots,X_k)$ is sampled according to $\cD$, then $X_i$ is distributed uniformly over $\Z_q$. We say that $f:\Z_q^k \to \{0,1\}$ supports one-wise independence if there is a one-wise independent distribution $\cD$ supported on a subset of the satisfying assignments of $f$, i.e., if $\veca\in\Z_q^k$ has positive probability under $\cD$ then $f(\veca)=1$. We say that $\cF$ {\em supports one-wise independence} if every function $f \in \cF$ supports one-wise independence. We say that $\cF$ {\em weakly supports one-wise independence} if there exists $\cF' \subseteq \cF$ such that $\rhomin(\cF')=\rhomin(\cF)$ and $\cF'$ supports one-wise independence.

\begin{theorem}[{\protect \cite[Theorem 2.17]{CGSV21-finite}}]\label{thm:one-wise}
If $\cF$ weakly supports one-wise independence, then $\cF$ is approximation-resistant to $o(\sqrt{n})$-space streaming algorithms.
\end{theorem}

Many natural families support one-wise independence. For readers familiar with some of these problems, we name some here without definitions: $\maxeksat$ for $k \geq 2$, $\mkor$, $\mcut$, $\maxqcol$, $\maxUG_q$ to name a few. All of these problems turn out to be approximation-resistant by the above theorem.

\paragraph*{Linear space lower bounds.}

Another direction of work has tried to extend the results of \cite{KK19}, i.e., $\Omega(n)$-space streaming lower bounds, to problems beyond $\mcut$. Here, we are far from a full understanding, but we do get approximation resistance for a (strict) subclass of families supporting one-wise independence. We define the families next.

For a function $f :\Z_q^k \to \{0,1\}$ and $\veca \in \Z_q^k$ we define the {\em width of $f$ at $\veca$} to be $\w_\veca(f) = \frac1q|\{\theta \in \Z_q | f(\veca + (\theta,\theta,\cdots,\theta))=1\}|$. We define the {\em width} of $f$ to be the quantity $\w(f) = \max_{\veca \in \Z_q^k}\{\w_{\veca}(f)\}$, i.e. the maximum over $\veca$ of the width of $f$ at $\veca$.
(Roughly the set $L_{\veca} = \{ \veca+(\theta,\theta,\cdots,\theta) | \theta \in \Z_q\}$ is a {\em line} through $\Z_q^k$ and $\w_\veca(f)$ measures the density of the intersection of this line with $f^{-1}(1)$, and the width of $f$ is the widest such intersection.)
We define the {\em width} of $\cF$, denoted $\w(\cF)$, to be the minimum over $f \in \cF$ of the width of $f$. Note that $1/q \leq \w(\cF) \leq 1$ for every $\cF$. Finally we say that $\cF$ is {\em wide} if $\w(\cF)=1$, i.e., the width is maximal. 

A simple example of a wide family is the $k$-equality function $f_{k\textsf{EQ}}$ where $f_{k\textsf{EQ}}(u_1,\ldots,u_k) = 1$ if and only $u_1 = \cdots = u_k$. Note that every wide family supports one-wise independence. But there exist functions supporting one-wise independence that are not wide: For example $\oplus_3:\Z_2^3 \to \{0,1\}$ given by $\oplus_3(a,b,c) = a + b + c \pmod 2$ supports one-wise independence but has width $1/2$.

The following theorem is shown in joint work with Chou, Golovnev, Velingker and Velusamy~\cite{CGSSV}.

\begin{theorem}[{\protect \cite[Theorem 1.1]{CGSSV}}]\label{thm:lin-ar} For every wide family $\cF$, $\maxF$ is approximation-resistant to $o(n)$-space streaming algorithms. 
\end{theorem}

We will not cover any aspects of the proof of this theorem in this article, except to say that it builds on the proof of \cite{KK19} following exactly the same sequence of steps, while replacing every step in their proof with ingredients needed to handle $k$-ary functions over non-Boolean alphabets. 
While the class of functions covered by this theorem is even smaller than the set covered by \cref{thm:one-wise}, it suffices to imply the following theorem which pins down the approximability of every $\maxF$ to within a factor of $q$. 

\begin{theorem}[{\protect \cite[Theorem 4.3]{CGSSV}}]  \label{thm:lin-coarse} For every family $\cF$ and every $\epsilon > 0$, $(\omega(\cF)-\epsilon,\rho(\cF)+\epsilon)$-$\maxF$ requires $\Omega(n)$ space for every streaming algorithm. Consequently, for every $\cF$ the largest $\alpha$ for which $\maxF$ is $\alpha$-approximable by a $o(n)$-space streaming algorithm satisfies $\alpha \in \left[\rhomin(\cF), \frac{\rhomin(\cF)}{\w(\cF)}\right]$. 
\end{theorem}

We remark that while \cref{thm:lin-coarse} immediately implies \cref{thm:lin-ar}, the proof in \cite{CGSSV} essentially derives the former from the latter using simple arguments.

\subsection{Aside: Ordering CSPs}

Before turning to some of the technical ingredients in the proofs we take a brief detour to cover an application of the results described above to a somewhat different class of optimization problems called ordering CSPs. We describe this class informally first: Recall that the solution space of the standard CSPs (the ones we work with in the rest of this paper) comes from a product set, namely an $n$-tuple of variables $(X_1,\ldots,X_n)$ takes values from $\Z_q \times \Z_q \times \cdots \times \Z_q$. A variation of this theme considers the setting where the variables need to be ordered, i.e., the $(X_1,\ldots,X_n)$ take on values from $\Sym_n = \{\pi:[n]\to[n] \mid \pi \textrm{ is one-to-one} \}$. (I.e., $X_i = \pi(i)$ where $\pi$ is a permutation.)
The natural notion of local constraints on ordering problems pick sequences of $k$ distinct variables out of the $n$ variables (as in standard CSPs) and look at the ordering from $\Sym_k$ induced by these $k$ variables and constrain them. Thus a constraint function in ordering CSPs is given by $\Pi:\Sym_k \to \{0,1\}$ and constraint families are a set of constraint functions. Thus for every $k$ and every family $\cF \subseteq \{\Pi: \Sym_k \to \{0,1\}\}$ we get an ordering CSP, denoted $\mocsp(\cF)$. (Note that unlike in standard CSPs, there is no notion of an alphabet or $q$ in the case of ordering CSPs.)

Two examples of ordering CSPs include the Maximum Acyclic Subgraph ($\mas$) problem and the Betweenness problem. The former asks, given a directed graph $G$, to find the largest acyclic subgraph in it. This problem is captured as $\mocsp(\{<\})$ where $< :\Sym_2 \to \{0,1\}$ satisfies $<(\pi)=1$ if and only if $\pi(1) < \pi(2)$. By placing the constraint $<(i,j)$ for every directed edge $(i,j)$ in a graph $G$, we get an $\mocsp(<)$ instance that exactly captures the $\mas$ instance. Betweenness is the ordering problem where constraints are given by a triple of variables and require that the ordering place the middle variable between the first and third (though allowing either of the first or the third to be the higher ranked variable). Once again it can be naturally formulated as an ordering CSP. 

With ordering CSPs again, one can ask what is the trivial approximability of an ordering CSP and when can a ordering CSP be solved non-trivially. Both questions turn out to have simple answers though somewhat disappointing ones from the algorithmic point of view. Note that a random ordering satisfies a constraint $\Pi$ with probability $\rho(\Pi) \eqdef \frac1{k!}\cdot |\Pi^{-1}(1)|$. Letting $\rho(\cF) = \min_{\Pi \in \cF}\{\rho(\Pi)\}$ we get that every instance of $\mocsp(\cF)$ has value at least $\rho = \rho(\cF)$, and thus $\rho$-approximation is trivial. It turns out that there are no algorithms (that can do better for any $\cF$ running in $o(n)$-space), as shown in the following theorem from joint work with Singer and Velusamy~\cite{SSV21}.

\begin{theorem}
For every $k$, every family $\cF\subseteq \{\Pi:\Sym_k \to \{0,1\}\}$ and every $\epsilon >0$, every streaming $(\rho(\cF)+\epsilon)$-approximation algorithm for $\mocsp(\cF)$ requires $\Omega(n)$ space.
\end{theorem}

\section{Some ideas behind the proofs}

\subsection{\texorpdfstring{The $\Omega(\sqrt{n})$ space lower bound for $\mcut$}{The sqrt(n) space lower bound for Max-CUT}}

We start with the lower bound from \cite{KKS15} on the $\mcut$ problem. We start with some basic ideas about lower bounds. Lower bounds in streaming are typically ``distributional''. To prove a lower bound on $\gbmF$ for some $\gamma,\beta,\cF$, for every sufficiently large $n$ we construct two distributions of instances on $n$ variables -- the $\yes$ and $\no$ distributions. The $\yes$ distributions are supported with probability $1-o(1)$ on instances from the set $\Gamma = \{\Psi | \val_\Psi \geq \gamma\}$. Similarly, the $\no$ distributions are supported with probability $1-o(1)$ on instances from the set $B = \{\Psi | \val_\Psi \leq \beta\}$. In the case of $\mcut$ we will thus consider a $\yes$ distribution supported (with probability one) on bipartite graphs, and $\no$ instances will have cut value at most $1/2 + o(1)$ (with probability $1-o(1)$). The goal is to prove that for any space $s$ algorithm $\ALG$ with $s = o(\sqrt{n})$ the distribution of the final state of $\ALG$ in the $\yes$ and $\no$ cases are very close in total variation distance. (For distributions $\cD,\cD'$ supported on some set $\Omega$ the total variation distance, denoted $\tvd{\cD-\cD'}$, is the quantity $\frac12 \sum_{\omega\in\Omega} |\cD(\omega)-\cD'(\omega)|$.)   Since the inputs are random, it suffices to consider deterministic $s(n)$-space bounded algorithms. 

Both distributions are parameterized by two constants: a small $\alpha \in (0,1)$ and large, but constant, integer $T$. The graphs are defined on vertex set $[n]$ and have roughly $(\alpha/2) \cdot T \cdot n$ edges. These edges come as the union of $T$ matchings $M'_1,\ldots,M'_T$, each with roughly $\alpha n/2$ edges.
In the $\no$ distribution these matchings will just be uniform matchings of the right size (we will get to the exact distribution of size shortly). In the $\yes$ distribution a random cut of $[n]$ is chosen by picking a vector $\vecx \in \{0,1\}^n$ uniformly at random and letting the cut be $\{i | x_i = 1\}$. The matchings $M_1,\ldots,M_t$ are uniform subject to the condition that every matched edge crosses the cut. The lower bound is proved by a ``hybrid argument'' involving $T$ steps. For $t \in \{0,\ldots,T\}$ let $S_t^Y$ denote the state of $\ALG$ after seeing the first $t$ matchings from the $\yes$ distribution, and similarly let $S_t^N$ denote the state of $\ALG$ after the first $t$ matchings from the $\no$ distribution. By definition we have $S_0^Y = S_0^N$. The key step is to prove that for every $t$, 
\begin{equation}
 \tvd{S_t^Y - S_t^N} \mbox{ is small assuming } \tvd{S_{t-1}^Y-S_{t-1}^N}\mbox{  is small,} \label{eq:maxcut-key}    
\end{equation}
and to use this result inductively to conclude $\tvd{S^Y_T-S^N_T}$ is small which shows that the two distributions are not distinguishable by small space algorithms. By construction the $\yes$ distribution is supported on bipartite graphs. If $\alpha T$ is sufficiently large then it can be argued by a standard Chernoff plus union bound that with probability $1-o(1)$, a graph from the $\no$ distribution also has value at most $1/2 + o(1)$ and together these suffice for the lower bound on $\mcut$. We thus turn to the proof of \cref{eq:maxcut-key}.

The upper bound works by designing two-party one way communication problem that captures the added distinguishability of $\yes$ from $\no$ conditioned on knowing $S_{t-1}^Y \approx_d S_{t-1}^N$ (where $\approx_d$ indicates that the two random variables are close in terms of total variation distance).
A rough abstraction of this problem is as follows: Alice, who knows $\vecx$ must send some information about it to Bob. This information may capture information such as $S_{t-1}^Y$ and/or $S_{t-1}^N$, both of which may in principle depend on $\vecx$, but should be limited to $o(\sqrt{n})$ bits. Now Bob, who gets to see $M_t$ which is either (in the $\yes$ case) a random matching crossing the cut given by $\vecx$ or (in the $\no$ case) a random matching, must distinguish the two. 

It turns out a problem very similar to this was already defined and studied in the literature. Specifically, Gavinksy, Kempe, Kerenedis, Raz and de Wolf~\cite{GKK+08} define the Boolean Hidden Matching (BHP) problem where Alice is given a uniform vector $\vecx \in \Z_2^n$ and Bob is given a  matching
$\tilde{M}$ with $m = \alpha n$ edges on vertex set $[n]$ drawn uniformly among all such matchings, and a $0/1$ labelling $\vecw\in\Z_2^m$ on the edges where in the $\yes$ case, the label $w_e$ of an edge $e = (i,j)$ satisfies $w_e = x_i + x_j \pmod 2$, while in the $\no$ case $w_e$'s are uniformly random and independent.
The goal of the communication is for Bob to distinguish the $\yes$ case from the $\no$ case. \cite{GKK+08} show that this problem requires $\Omega(\sqrt{n})$ bits of communication to achieve constant advantage in distinguishing. (The advantage of a protocol is the probability that the protocol outputs $1$ in the $\yes$ distribution minus the probability it does so in the $\no$ distribution. Specifically the \cite{GKK+08} result shows that for every $\delta > 0$ there exists $\tau > 0$ such that for every $\alpha < 1/2$ and every sufficiently large $n$, every protocol that achieves advantage $\delta$ must communicate at least $\tau\sqrt{n}$ bits. These quantifiers are somewhat important as we will see below.)

To use the BHM lower bound from \cite{GKK+08} we need to address two issues. First the input to Bob in the BHM problem is not the same as coming from the streaming problem. This problem is easy to deal with --- Bob is getting more information in the BHM problem than in the motivating $\mcut$ based problem, and this only makes the lower bound even stronger. Formally Bob can reduce an instance of BHM to the streaming inspired-problem by dropping all the edges $e$ that have label $w_e = 0$. This gives Bob roughly $\alpha n/2$ edges (since each edge crosses the cut with probability roughly $1/2$ and these are roughly independent events) reducing exactly to the setting in the streaming-inspired problem. 

The second and more important issue is that the BHM problem was only ``roughly'' motivated by the streaming problem above --- we need a more careful and formal argument connecting the two. Formally we consider the random variables, $S^Y_t$, $S^N_t$ and a hybrid variable $\tilde{S}$, where $\tilde{S}$ is the state of $\ALG$ on receiving $M_1,\ldots,M_{t-1}$ from the $\yes$ distribution and $M_t$ from the $\no$ distribution. The BHM lower bound immediately implies that $\tilde{S} \approx_d S^Y_t$: The only difference between the two states is the $t$th input which comes from the $\yes$ distribution for $S^Y_t$ and from the $\no$ distribution for $\tilde{S}$; and the setup of BHM allows Alice to generate and communicate $S^Y_{t-1}$ to Bob allowing Bob to compute the final state and use $\ALG$ to distinguish them. To complement we also have $\tvd{\tilde{S}-S^N_t} \leq \tvd{S^Y_{t-1}-S^N_{t-1}}$ by the data processing inequality: $\tilde{S}$ is determined by $S^Y_{t-1}$ and $M_t \sim \no$ while $S^N_t$ is determined from $S^N_{t-1}$ and $M_t$. We stress a subtle point here: It is crucial that $M_t$ is independent of $S^Y_{t-1}$ and $S^N_{t-1}$ for this inequality to be applicable, and this does hold in our case since the $\no$ distribution is independent of $\vecx$ which is the only variable connecting the different matchings in the $\yes$ case. (This subtlety is the reason why extensions of this proof apply only to families supporting one-wise independence, or only give sketching lower bounds.)

We also comment briefly on the choice of various parameters such as $\alpha$, $T$, $\epsilon$ (where our goal is to prove hardness of $(1,1/2+\epsilon)$-$\mcut$), $\delta$ (the advantage allowed in BHM) and $\tau$ (where the space lower bound is $\tau\sqrt{n}$). We want our bound to hold for every $\epsilon > 0$ so given 
$\epsilon$, we first pick $\alpha$ small enough for the BHM lower bound to hold. In our case it holds for every $\alpha < 1/2$. Given this choice of $\alpha$ we pick $T$ large enough so that a graph from the $\no$ distribution with $\alpha T n$ edges is very likely not to have a $\mcut$ of fractional size more that $1/2+\epsilon$. Given this choice we pick $\delta$ small enough so that $T$ applications of the hybrid argument still lead to negligible advantage in distinguishing $\yes$ from $\no$. Finally the $\tau$ we obtain is whatever is guaranteed by the BHM lower bound for this choice of $\delta$. 

Our eventual streaming and sketching lower bounds will extend the ideas from above, but we will return to those after describing the algorithms for $\mdcut$ from \cite{GVV17,CGV20}. 

\subsection{\texorpdfstring{Bias-based algorithms for $\mdcut$}{Bias-based algorithms for Max-DICUT}}

The key ingredient in the algorithm of \cite{GVV17} for $\mdcut$ is the notion of the ``bias'' of a graph on vertex set $[n]$. For a vertex $v$ in a directed graph, let
$\indeg(v)$ denote the number of incoming edges into $v$ and let $\outdeg(v)$ denote the number of outgoing edges. Now define $\bias(v) = \indeg(v)-\outdeg(v)$, and define $\bias(G) = \frac1{2m} \sum_{v\in [n]} |\bias(v)|$. Thus if we term the vector $(\bias(v))_{v\in[n]}\in \R^n$ to be the bias-vector of the graph, then the bias of the graph is essentially the $\ell_1$ norm of this vector up to normalization. As mentioned already, the $\ell_1$-norm and hence the bias of a graph can be estimated arbitrarily well by a streaming algorithm presented with a stream of edges using an algorithm from \cite{Ind06}. The key to the algorithms of \cite{GVV17} and \cite{CGV20} are inequalities relating the bias of a graph to the dicut value, that allow them to output lower bounds of the value of the dicut. (For uniformity we will only talk about the fractional value here and later and use $\val_G$ to denote this quantity.)

Note that by definition $0 \leq \bias(G) \leq 1$ for every graph $G$ on $m$ vertices. \cite{GVV17} show that $\val_G \leq \frac{1+\bias(G)}{2}$. This inequality follows easily from the observation that every cut must leave at least $|\indeg(v)-\outdeg(v)|$ of the edges incident to $v$ uncut. Since every uncut edge may be counted twice by this process, we get a lower bound of $\frac12\sum_v |\indeg(v)-\outdeg(v)|$ on the number of uncut edges.

\cite{GVV17} complement upper bound above with a lower bound: For every $G$ we must have $\val_G \geq \bias(G)$. This is ``constructive'' (though not in streaming sublinear space) --- the greedy cut which puts all vertices with positive bias on the sink side of the cut and the rest on the source side achieves this. (A simple argument to see is iterative: Remove directed cycles from the graph one at a time till we get a DAG. This does not alter the bias. Now remove maximal length directed paths - each such path contributes one to the non-normalized bias, and also contributes at least one edge to the dicut since by maximality the source of the path must have zero indegree and the sink must have zero out degree.)

Combining the two bounds above with the lower bound $\val_G \geq 1/4$ for every $G$ gives a $.4$ approximation algorithm: The algorithm computes $\bias(G)$ and outputs $\max\{\bias(G),1/4\}$. To improve on this \cite{CGV20} give an improved lower bound on $\val_G$ when $\bias(G) \leq 1/3$. Their bound is also ``constructive'' - they consider a random dicut where each vertex of positive bias is placed on the sink side with probability $1/2+\delta$ independently (for some parameter $\delta$ that we will optimize later). Remaining vertices are placed on the sink side with probability $1/2-\delta$ independently. They analyze the cut produced by this rounding after optimizing over $\delta$ and use the expected size as an additional lower bound. We won't reproduce their bound or analysis here, but only comment that the analysis involves optimizing degree two rational functions in $\delta$. This already gives them a 
$4/9$ approximation algorithm.

The choice of a single rounding probability for all vertices in the graph is somewhat surprising. (This probability may depend on the graph and bias, but once the graph is fixed all vertices get rounded with the same probability.) It seems like a choice made for ease of analysis - optimizing a single variable $\delta$ is easier than optimizing $n$ variables! 
One could nevertheless ask --- could we have done better with more careful choices?
The surprising result from \cite{CGV20} is that this won't help and indeed no $o(\sqrt{n})$-space algorithm can improve on the bound above! So somehow $\bias(G)$ is the right quantity to compute, and rounding independently with the same probability for all vertices (upto the choice of the preferred side) is the right algorithm! 

\subsection{The framework of \cite{CGSV21-finite}}

To extend the algorithm of the previous section to problems beyond $\mdcut$, we need to understand what are notions of bias of a variable and of the whole instance for $\maxF$ for general $\cF$. (In this discussion we will assume $\cF$ has a single function $f$ though extensions to more functions are straightforward.) Recall that in the $\mdcut$ problem constraints arrive as pairs $(i,j)$ where the edge goes from vertex $i$ to vertex $j$. Thus the in-degree of a vertex could be abstract as the number of constraints in which it is the second variable, and out-degree as the number of constraints where it is the first variable. We use this to motivate a new notion of bias of a variable $X_i$, denoted $\dbias(i)$ (for detailed bias): This will be a $k$ dimensional vector whose $j$th coordinate $\dbias(i)_j$ records the number of constraints in which  $X_i$ appears as the $j$th variable in a constraint. 
Considering all the biases of all vertices gives us an $n\times k$ matrix $B=B(\Psi)$ with $B(i,j) = \dbias(i)_j$ that ``represents'' an instance $\Psi$.

The main idea in \cite{CGSV21-finite} can roughly be captured as follows: 
\ifnum\lipics=1
\emph{If there exists $t \in \N$ and instances $\Psi_g$ and $\Psi_b$ on $t$ variables with $B(\Psi_g) = B(\Psi_b)$ such that 
$\val_{\Psi_g} \geq \gamma$ and $\val_{\Psi_b} \leq \beta$ then $\maxF$ can not be solved in $o(\sqrt{n})$ space by a sketching algorithm. Else it can be solved by a polylogarithmic space linear sketching algorithm.}
\else
\begin{quote}
\emph{If there exists $t \in \N$ and instances $\Psi_g$ and $\Psi_b$ on $t$ variables with $B(\Psi_g) = B(\Psi_b)$ such that 
$\val_{\Psi_g} \geq \gamma$ and $\val_{\Psi_b} \leq \beta$ then $\maxF$ can not be solved in $o(\sqrt{n})$ space by a sketching algorithm. Else it can be solved by a polylogarithmic space linear sketching algorithm.  }
\end{quote}
\fi
A priori neither statement should be obvious and we will give some idea below as to why they are true. Furthermore even if the statements are true it is not clear how to decide which of the two conditions hold (since a priori one may have to enumerate over all $n$ and all pairs of instances to determine if the condition is true). It turns out all the issues get answered rather nicely jointly. It turns out that it suffices to consider (weighted) instances on $kq$ variables to answer the final question, and studying the space of these instances also leads to the algorithms and the lower bounds. 

Below we elaborate on this and in particular why it suffices to consider instances on a finite number of variables. We work with the simpler setting of Boolean CSPs (so $q=2$) on literals, i.e., when constraints can be applied to variables as well as their negations.
We note that the resulting setting ($|\cF|=1$, $q=2$ and constraints on literals) is the case considered in a preliminary work~\cite{CGSV21-boolean}, whereas the more general result comes from \cite{CGSV21-finite}. While the latter is a stronger result, the former offers more intuition into the proofs.

In this setting of constraints over Boolean literals, we show we only need to consider instances involving $k$ variables --- and we explain how this happens. 
Suppose there are two instances on $n$ variables: $\Psi_g$ with $\val_{\Psi_g} \geq \gamma$ and $\Psi_b$ with $\val_{\Psi_b} \leq \beta$ satisfying $B(\Psi_g) = B(\Psi_b)$.
We show how to simplify the two CSPs. From now onwards it will be convenient to think of a weighted CSP instance as being a distribution on constraints - where a constraint is chosen with probability proportional to its weight. Now, since we are considering Boolean CSPs on {\em literals} we can flip variables as necessary (by flipping literals in all constraints) till we get that $1^n$ is the assignment achieving $\val_{\Psi_g}(1^n) \geq \gamma$. To preserve $B(\Psi_g) = B(\Psi_b)$ we flip variables in $\Psi_g$ and $\Psi_b$ together. Note that this flipping preserves $\val_{\Psi_b} \leq \beta$. Next we observe that we can assume $\Psi_g$ and $\Psi_b$ are symmetric under permutations: I.e. if some constraint
$C(X_1,\ldots,C_n)$ appears in $\Psi_g$ with some probability $p$ then for every permutation $\pi:[n]\to[n]$ the constraint $C(X_{\pi(1)},\ldots,X_{\pi(n)})$ also appears in $\Psi_g$ with the same probability $p$. (We can convert any $\Psi$ to a $\Psi'$ satisfying this feature as follows: To pick a random constraint of $\Psi'$, pick a random constraint $C$ of $\Psi$ and a uniformly random permutation and let the constraint produced by $C(X_{\pi(1)},\ldots,X_{\pi(n)})$.) This transformation preserves $\val_{\Psi_g} \geq \gamma$ since $1^n$, the assignment achieving the maximum value is fixed under permutations. We also have that $\Psi_b$ is closed under permutations since the (empty!) set of assignments that achieves value greater than $\beta$ is also closed under permutations.
The fact that $\Psi_g$ and $\Psi_b$ are symmetric under permutations make them very simple: All that determines these instances is the distribution supported on $\Z_2^k$ indicating the pattern of negations of the $k$ variables in a randomly chosen constraint. The names of the variables are no longer relevant --- since they are just a uniformly random sequence of $k$ distinct variables! Suppose $\cD_Y$ represents the distribution on $\Z_2^k$ given by $\Psi_g$ and $\cD_N$ the distribution given by $\Psi_b$. We now study these distributions further and they will lead us to the answers to the three issues raised earlier.

\snote{These are typeset weirdly: Labelled 5.3.0.1 (parent is 5.3)}

\paragraph*{The space of $\cD_Y$ and $\cD_N$.} We can go back from distributions $\cD$ over $\Z_2^k$ to instances $\Psi_{\cD}$ of $\maxF$ on $k$ variables $X_1,\ldots,X_k$ as follows: A random constraint of $\Psi_{\cD}$ is of the form $f(X_1 \oplus b_1,\ldots,X_k \oplus b_k)$ where $\vecb = (b_1,\ldots,b_k) \sim \cD$. Now the fact that $\cD_Y$ came from an instance $\Psi_g$ of value at least $\gamma$ implies that the all $1$'s assignment satisfies $\Psi_{\cD_Y}$. The fact that $B(\Psi_g) = B(\Psi_b)$ implies that $\cD_Y$ and $\cD_N$ have the same marginals. It remains to interpret the implication that $\Psi_b \leq \beta$: We stress that it does not mean $\Psi_{\cD_N}$ has value less than $\beta$ - indeed $\Psi_{\cD_N}$ can have value much larger than that or even $\gamma$! The implication turns out to be exactly the following: ``For every $p \in [0,1]$ if $X_1,\ldots,X_k$ are assigned values identically and independently according to $\bern(p)$ (i.e., they take values in $\Z_2$ with $\Pr[X_i = 1] = p$),  then the expected value $\val_{\Psi_{\cD_N}}(X_1,\ldots,X_k) \leq \beta$.'' I.e., no identical and independent probabilistic assignment to the variables satisfies many constraints. 

It turn out we can now capture these considerations on $\cD_Y$ and $\cD_N$ in a nice mathematical framework and that will lead to matching algorithms and lower bounds. Note that a distribution on $\Z_2^k$ can be viewed as a vector in $\R^{2^k}$ in a natural way, and the space of all distributions is a convex set in $\R^{2^k}$. Now let $\sgyf$ denote the subset of this set representing distributions $\cD$ such that $\val_{\Psi_{\cD}}(1^k) \geq \gamma$.
Similarly let $\sbnf$ denote the subset of distributions $\cD$ such that for every $p \in [0,1]$, $\Exp_{\vecb \in \bern(p)^k} \left[ \val_{\Psi_{\cD}}(\vecb)\right]\leq \beta$. Both these sets are convex sets! (In particular for every $p$, the constraint $\Exp_{\vecb \in \bern(p)^k} \left[ \val_{\Psi_{\cD}}(\vecb)\right]\leq \beta$ is a linear constraint on $\cD$, though we have infinitely many such constraints.) By construction the two sets are disjoint for $\beta < \gamma$, but they may still contain distributions with matching marginals! To see this we may project these two sets to their marginals: So let $\kgyf \subseteq \R^k$ be the set of marginals of all distributions in $\sgyf$ and similarly let $\kbnf$ be the marginals of $\sbnf$. The discussion thus far has reduced the question: ``Do there exist $n$ and instances $\Psi_1$ and $\Psi_2$ on $n$ variables with $B(\Psi_1) = B(\Psi_2)$ such that 
$\val_{\Psi_1} \geq \gamma$ and $\val_{\Psi_2} \leq \beta$?'' to the much simpler and finite dimensional question ``Do $\kgyf$ and $\kbnf$ intersect?''. (An affirmative answer to one question implies an affirmative answer to the other.)

Before turning to show why this leads to algorithms or lower bounds we first point out that the question of the intersection of these two sets is decidable. Specifically the intersection question can be posed as polynomial inequalities in $2^k+1$ variables ($2^k$ from $\cD$ and one from $p$) of degree at most $k+1$ with one variable ($p$) being universally quantified and the rest being existentially quantified. Results in the quantified theory of reals~\cite{BasuPR} easily show how to decide this question in space polynomial in the input size, which in our case is roughly $2^k$ to represent the function $f$ (and whatever else is needed to specify $\gamma$ and $\beta$). 

\paragraph*{Sketching lower bound when $\kgyf \cap \kbnf \ne \emptyset$.} It turns out that the existence of two distributions with matching marginals is the crux of the $\mcut$ lower bound of \cite{KKS15} and so extending to other settings is a reasonable hope. Specifically the $\mcut$ lower bound relies on
$\cD_Y = \unif(\{00,11\})$ and $\cD_N = \unif(\Z_2^2)$. To extend to other problems and distributions, we use the same approach of dividing a long stream of constraints into $T$ substreams of length $\alpha n$. A communication problem captures the additional information gained by a substream while a hybrid argument combines the information gained from the substreams. Both steps turn out to be different though and we elaborate on them below.

The BHM problem could be interpreted as arising from the associated distributions above in two different ways. In both Alice gets $\vecx \in \Z_2^n$ and Bob's first input is a matching on $[n]$, which specifies potential constraints: Bob's second input can be interpreted in two ways: (1) For each constraint, he gets information on whether $\vecx$ satisfies the constraint or not, (2) Using the fact that $\cD_Y$ is uniform on a subgroup of $\Z_2^2$, Bob gets input on which coset the variables in the constraint come from. The first interpretation doesn't seem naturally amenable to the lower bound techniques which seem more tailored to understanding inputs that are uniform in the $\no$ case. The second interpretation seems restricted to groups and cosets and in particular does not seem to support $\cD_Y$ not being uniform on a set, leave alone a subgroup. However it is possible to extend this approach beyond such algebraic settings and this is what is done in \cite{CGSV21-finite}. To do so they introduce the $(\cD_Y,\cD_N)$-Randomized Mask Detection Problem (RMD) which is again a distribution distinguishability problem in the one-way communication setting: Here Alice gets a vector $\vecx \in \Z_2^n$ and Bob gets a $k$ hypermatching with $m = \alpha n$ edges. Additionally Bob gets a vector $\vecw \in \Z_2^{km}$, or equivalently, one vector in $\Z_2^k$ associated with each hyperedge of the matching. In the $\yes$ case this vector associated with a hyperedge  is the labels of $\vecx$ restricted to the vertices incident to the hyperedge masked (i.e., xor-ed, or summed in $\Z_2$) by a vector $\vecb \in \Z_2^k$ drawn according to $\cD_Y$. Each mask vector $\vecb$ is drawn independently for every hyperedge. The $\no$ distribution is similar with the difference that now $\vecb\sim \cD_N$ independently for each edge. 

\cite{CGSV21-finite} give a $\Omega(\sqrt{n})$ communication lower bound for this problem to achieve any constant advantage (this time for $\alpha < 1/k$). The lower bound works in two parts. First the extend the proof from \cite{GKK+08} to general $k$ in the setting where $\cD_N$ is uniform on $\Z_2^k$. (As noted above this setting seems amenable to their proof technique). The second part of the proof shows how to use the first part to show hardness of RMD on distributions $\cD_1$ and $\cD_2$ that differ in a ``simple'' way (in particular they differ in probabilities of at most four structured points in their support). 
They then complement this by showing that one can move from every $\cD_Y$ to every $\cD_N$ (with matching marginals) using a finite number of steps (as a function of $q$ and $k$) where each step creates a ``simple'' difference in the sense above. A series of triangle inequalities now shows that $(\cD_Y,\cD_N)$-RMD is also indistinguishable to $o(\sqrt{n})$-communication protocols.

To convert the RMD lower bound into a lower bound on $\maxF$ we first need to interpret the RMD inputs as constraints of a $\maxF$ problem, and then to prove that combining $T$ substreams preserves indistinguishability by streaming algorithms. The first step is natural: We apply constraints so that the hidden vector $\vecx$ is expected to satisfy $\gamma$ fraction of the constraints in the $\yes$ case: Specifically if a hyperedge gives Bob the information $\vecx|_S + \vecb$ corresponding to the restriction of $\vecx$ to some sequence $S$ of $k$ variables masked by $\vecb$, then the resulting constraint negates literals according to $\vecx|_S + \vecb$, so that after the negations are applied, the input to the constraint is $\vecb$ which, by the condition that $\cD_Y \in \sgyf$ is expected to satisfy the constraint with probability $\gamma$. Similarly in the $\no$ case every assignment is expected to satsify the constraint with probability at most $\beta$. Taking sufficiently many constraints (i.e., $\alpha T \to \infty$) allows us to apply Chernoff bounds and the union bound to conclude tight bounds on the value of the resulting CSPs in the $\yes$ and $\no$ case. 

The tricky part turns out to be the combination. When $\cD_N$ is uniform, the same hybrid argument as in the $\mcut$ case works and using this twice we conclude that if $\cD_Y$ and $\cD_N$ have uniform marginals then the resulting CSP instances are indistinguishable by $o(\sqrt{n})$ space streaming algorithms. \cite{CGSV21-finite} also cover some slight extensions that allow them to cover hardness of $\mdcut$ where the underlying distributions do not have uniform marginals. But for general $\cD_Y$ and $\cD_N$ with non-uniform marginals the method truly breaks down and produces instances of $\maxf$ that are distinguishable by polylogspace algorithms as pointed out by~\cite{CKP+21}. However in such cases it is possible to show that no sketching algorithm can work. This relies on an easy reduction from RMD to a $T$-player simultaneous communication problem where $T$ players each get inputs independently according to the distribution of Bob's input and then need to communicate short messages to a referee whose goal is to distinguish the inputs being all from the $\yes$ distribution or from the $\no$ distribution. This yields the lower bound of \cref{thm:cgsv}.

\paragraph*{A sketching algorithm when $\kgyf \cap \kbnf = \emptyset$.} We now turn to the complementary result, giving a sketching algorithm  when $\kgyf \cap \kbnf = \emptyset$. If the sets do not intersect then there must be a hyperplane in $\R^k$ separating them. Let this plane be given by $\lambda_1,\ldots,\lambda_k$ and $\tau$ so that 
$\{\veca \in \R^k | \sum_{i\in[k]} \lambda_i a_i \geq \tau\}$ contains $\kgyf$. Since $\kgyf$ and $\kbnf$ are closed sets if they are disjoint there must be a gap separating them and so we also have $\theta > 0$ so that $\{\veca \in \R^k | \sum_{i\in[k]} \lambda_i a_i \leq \tau-\theta\}$ contains $\kbnf$.

It is natural to think of $\lambda_i$ as representing a preference that the $i$th variable in this constraint has for taking the value $1$ with higher $\lambda_i$'s representing higher preferences. When the $i$th variable in a constraint is negated we let $- \lambda_i$ capture its preference. These preferences allow aggregation across constraints and yield the definition: For $j \in [n]$, 
let $\bias(\Psi,j)$ be the sum of the appropriate $\lambda$ values over all constraints that variable $j$ participates in. Define $\bias(\Psi) = \frac1m \sum_{j=1}^n |\bias(\Psi,j)|$. (We note that these definitions extend the $\mdcut$ notions exactly). $\bias(\Psi)$ can be estimated as previously by appealing to $\ell_1$ norm estimation algorithms. The algorithm for $\gbmF$ now reports $\yes$ if and only if $\bias(\Psi) \geq \tau - \theta/2$. 
This algorithm turns out to be correct, using analysis ideas that follow in a straightforward way from the construction of the convex sets. In case the reader wonders where the $\ell_1$ estimator is suggested in the construction of the convex sets, this happens in the step where we passed from a general 
$\Psi_1$ and $\Psi_2$ with matching detailed bias matrix $B$, to assuming $1^n$ achieves the maximum value of $\Psi_1$. The computational challenge behind this vertex is to compute the maximal satisfying assignment, flipping literals of $\Psi_1$ according to this, and then computing its bias. This step is achieved computationally by the $\ell_1$ norm estimation algorithm!

\paragraph*{The general case.}

Up to now we focussed on the simpler case of $\cF = \{f\}$, $q=2$ and constraints being applied to literals. It turns out that this is the exact setting considered in the early version \cite{CGSV21-boolean}. The extension to 
the general case, where $\cF$ is not a singleton, constraints are applied only to variables, and $q$ is general, appears in \cite{CGSV21-finite}. The extensions do manage to work out with no surprises (at least no unpleasant ones).

The elimination of the need to work with literals is the conceptually hard step but works out by working with $qk$ variables which is different from $k$ even in the Boolean case. Roughly our simplified picture of the extremal examples $\Psi_1$ and $\Psi_2$, uses two variables for each of the $k$ positions in constraint that a variable can appear in: one corresponding to the unnegated variable $X$ and one to the negated variable. To extend to general $q$ we now use $q$ variables per coordinate $i \in [k]$ --- with variable $X_{i,\sigma}$ roughly capturing the ``literal'' $X_i + \sigma \pmod q$. 
The extension to larger sets $\cF$ is simple, we augment the detailed-bias information as well as the marginals to include information about which function $f \in \cF$ the constraint is working with. This leads to sets $\sgyf$, $\sbnf$ that are extended to capture distributions over $\cF \times \Z_q^k$ and the marginals
$\kgyf$, $\kbnf$ now project to $\cF \times [k] \times \Z_q$ dimensions. Somewhat surprisingly both the algorithm and the lower bounds extend to this setting with the $\ell_1$ norm estimator replaced by an $\|\cdot\|_{1,\infty}$-norm estimator\footnote{For matrix $M$ with rows indexed by $[n]$ and columns by $\Z_q$ the $(1,\infty)$-norm is given by $\|M\|_{1,\infty} = \sum_{i=1}^n \max_{j\in \Z_q} |M_{ij}|$.} of \cite{AKO11}. Deciding intersection of the two sets reduces to quantified systems with $2$ alternations, and roughly $|\cF|q^k$ variables and degree $k$. Perhaps the most complex part of the extension is the extension of the streaming lower bounds which work with two variants of the RMD problem. Also the reduction of the communication problems to the streaming problems is a bit delicate due to the absence of literals but works out in the end. We omit the many details, referring the reader to the original paper~\cite{CGSV21-finite} for those.

\section{Future directions}

One can hope for many possible extensions to the dichotomy reported in \cref{thm:cgsv}. Perhaps the dichotomy extends as is to streaming algorithms (i.e., beyond sketching), perhaps even for linear space, perhaps even for randomly ordered streams, and maybe even for multipass algorithms. Unfortunately, while several extensions are still possible, the clean dichotomy does seem to fray quite a bit for each possible extension. 

At the moment it is still plausible that the dichotomy extends as is to all $o(\sqrt{n})$ space streaming algorithms though there is no strong evidence in either direction. For space beyond $\sqrt{n}$ there do seem to be a number of new candidate algorithms so our expectation would be that the current dichotomy won't hold and there may exist more than two classes of problems. We note that we don't have concrete theorems proving this though. For randomly ordered streams as well as multipass algorithms there seem to be new algorithms in polylogarithmic space. This is the subject of an upcoming work by the author with Saxena, Singer and Velusamy. Finally the multipass setting seems to be the most challenging for the lower bounds. Here some remarkable works~\cite{AKSY20,AssadiN}, have shown strong space lower bounds but for progressively weak approximations. Here an interesting challenge is to establish a tight lower bound for any non-trivial CSP for arbitrarily large (but constant) number of passes. 

\ifnum\lipics=0
\section*{Acknowledgements} \acktext
\fi 

\ifnum\lipics=0 \bibliographystyle{alpha}

\newcommand{\etalchar}[1]{$^{#1}$}

\else  
\bibliography{lipics-main}
\fi 

\end{document}